\documentclass[pra,aps,superscriptaddress,twocolumn,longbibliography]{revtex4-1}

\setcitestyle{super}

\usepackage{times}
\usepackage{graphicx,subfigure}
\usepackage{bm}
\usepackage{amsmath,amsfonts,amsthm,amssymb}
\usepackage{mathtools}
\usepackage[abs]{overpic}
\usepackage{color}
\usepackage[usenames,dvipsnames]{xcolor}
\usepackage{soul}
\usepackage{comment}
\usepackage[breaklinks]{hyperref}
\usepackage{epstopdf, epsfig}
\usepackage{grfext,grffile}
\usepackage[normalem]{ulem}

\newcommand{\bu}{{\bf u}}
\newcommand{\bR}{{\bf R}}

\newcommand{\vn}[1]{\left\| #1\right\|}  

\newcommand{\corr}{\color{black}}
\newcommand{\rroc}{\color{black}}

\begin{document}


\title{From  the butterfly effect to intrinsic randomness:\\ the spontaneous growth of singular shear flows}
\author{Simon Thalabard}\email{simon.thalabard@ens-lyon.org}\affiliation{Instituto Nacional de Matem\'atica Pura e Aplicada -- IMPA, 22460-320 Rio de Janeiro, Brazil}
\author{J\'er\'emie Bec}\email{jeremie.bec@mines-paristech.fr}\affiliation{MINES ParisTech, PSL Research University, CNRS, CEMEF, Sophia--Antipolis, France}
\author{Alexei Mailybaev }\email{alexei@impa.br}\affiliation{Instituto Nacional de Matem\'atica Pura e Aplicada -- IMPA, 22460-320 Rio de Janeiro, Brazil}

\begin{abstract}
The butterfly effect is today commonly identified with the sensitive dependence of deterministic chaotic systems upon initial conditions.  However, this is only one facet of the notion of unpredictability pioneered by Lorenz, who actually predicted that multiscale fluid flows could spontaneously lose their deterministic nature and become intrinsically random. This effect, which is radically different from chaos, have remained out of reach for detailed physical observations. Here, we substantiate this scenario by showing that it is inherent to the elementary Kelvin--Helmholtz hydrodynamical instability of an initially singular shear layer. We moreover provide evidence that the resulting macroscopic flow  displays universal statistical properties that are triggered by, but independent of specific physical properties at micro-scales. This spontaneous stochasticity is interpreted as an Eulerian counterpart to  Richardson's relative dispersion of Lagrangian particles, giving substance to the intrinsic nature of randomness in turbulence.
\end{abstract}

\maketitle

\noindent In his seminal works on unpredictability of atmospheric motion, Edward N.~Lorenz~\cite{lorenz1963deterministic,lorenz1969predictability} formulated a visionary conjecture on the sensitive dependence of deterministic systems upon initial perturbations or errors.  He distinguished between two kinds of unstable behaviors: Either the mismatch between two replica of the same system can be made arbitrarily small by sufficiently reducing their initial discrepancy or, alternatively, the two systems reach diverging states, no matter how small they initially differ from each other. 
The first type of behavior arises in  chaotic systems, wherein small perturbations are amplified exponentially in time.
 This phenomenon is now celebrated as the \emph{butterfly effect}~\cite{gleick2011chaos} and has been widely used to link unpredictability and complexity~\cite{boffetta2002predictability}.

The second behavior corresponds to a much more drastic \emph{intrinsic} materialization of unpredictability. 
 In Lorenz's own words: \textit{certain formally deterministic fluid systems which possess many scales of motion are observationally indistinguishable from indeterministic systems; specifically, that two states of the system differing initially by a small “observational error” will evolve into two states differing as greatly as randomly chosen states of the system within a finite time interval, which cannot be lengthened by reducing the amplitude of the initial error}~\cite{lorenz1969predictability}.  According to Lorenz,  this second type of unpredictability  pertains to fluid flows with a sufficient amount of kinetic energy at small scales, meaning that  the velocity field should have a singular spatial structure.
As pointed out by Palmer \textit{et al.}~\cite{palmer2014real}, the dynamical formulation of such fluid systems must be ill-posed: solutions do not depend continuously on initial conditions, therefore, allowing a finite-time separation between initially undistinguishable systems. 
In spite of being supported by phenomenological arguments\cite{leith1972predictability,ruelle1979microscopic,eyink1996turbulence,palmer2000predicting,boffetta2001predictability,mailybaev2016spontaneously,mailybaev2017toward,biferale2018rayleigh}, a clear physical evidence is still lacking whether such a behavior indeed arises in genuine fluid flows, hereby making them infinitely less predictable than chaos.

In this work, we demonstrate that this scenario is in fact relevant for the fundamental instability in classical fluid mechanics: the Kelvin-Helmholtz (KH) instability which describes the  growth of a shear layer from an initially discontinuous velocity profile\cite{drazin2002introduction}. This instability is an important constituent in dynamical phenomena~\cite{Matsuoka:2014} ranging from the micro-world of quantum fluids~\cite{blaauwgeers2002shear,takeuchi2010quantum} to macroscopic motions in Earth’s atmosphere and oceans~\cite{smyth2012ocean}, extending further to astrophysical problems of supernovae~\cite{wang2001instabilities} and interstellar clouds~\cite{vietri1997survival}.
While formally deterministic, the singular shear layer problem is however ill-posed in ideal fluid mechanics~\cite{szekelyhidi_weak_2010}.
The dominant viewpoint generally attributed to the French mathematician J.~Hadamard is that such problems are physically meaningless, unless suitably regularized, so that solutions become  uniquely determined and exhibit continuous dependence upon a class of initial data \cite{hadamard1902problemes,sobolev2016partial}. This viewpoint has motivated numerous studies employing  analytic perturbation of the vortex sheet initial datum, that carefully analyze finite-time solutions of such  deterministic and well-posed problems \cite{caflisch1989singularity,duchon1988global,cowley1999formation,sulem1981finite};
\corr
it also  motivates the use of KH instability as a test-bed numerical problem  \cite{lecoanet2015validated},  and suggests the deterministic approach to unveil the physics of shear layer flows \cite{corcos1981deterministic,corcos1984mixing}
\rroc

The Hadamard viewpoint surely appears quite at odds with  Lorenz's intuition.
\corr
It also fails to account for the essential footprint of KH flows:  their   visually striking large-scale features, that allow for identifying easily occurrences of KH instabilities across the physical communities, regardless of the detailed triggering mechanisms.  This is the fundamental and long-standing observation that  certain macroscopic statistical  features of  freely evolving shear flows are apparently only mildly dependent upon initial conditions \cite{sommeria1991final,suryanarayanan2014free}.
\rroc
Our objective here is to point out that the  classical Kelvin-Helmholtz singular shear layer becomes in fact well-posed and physically relevant when formulated in a probabilistic sense.
\corr
This implies a change from a deterministic to a probabilistic viewpoint, in order to recouncile Hadamard's to Lorenz's.
\rroc 

 To unveil the intrinsically random nature \corr of the shear layer dynamics\rroc, we employ a  probabilistic approach inspired from earlier studies connected to the theory of \emph{spontaneous stochasticity}\cite{eijnden2000generalized,falkovich2001particles,gawedzki2001turbulent,le2002integration,kupiainen2003nondeterministic,eyink2013flux} describing unpredictability of trajectories of Lagrangian particles advected by rough (non-smooth) deterministic velocity fields,  a key notion beneath  the modern view on turbulent mixing\cite{bernard1998slow,shraiman2000scalar,iyer2018steep,dombre2018zero,drivas2017lagrangian}. This approach makes use of micro-scale regularizing mechanisms, such as viscous dissipation and thermal fluctuations, to select relevant physical solutions at the macro-scale. 
Inferring from this connection, our main result is to reveal the intrinsic space-time randomness of the shear layer growth, characterized by the existence of a \textit{universal} non-trivial probability distribution of the velocity field. Therefore, despite unpredictability, the resulting ideal flows possess well-defined statistical properties at \textit{finite times}, which are triggered by, but not sensitive to the nature of infinitesimal micro-scale details.
 
\bigskip
\noindent\textbf{Results}\\
\noindent\textbf{\small The singular shear layer.}  The classical mathematical formulation of the KH instability refers to perturbations of the idealized interface between two parallel streams with the velocity difference $U$. We will focus on two-dimensional incompressible formulation, as it is more accessible for accurate numerical simulations; in applications, this formulation refers to large-scale motions in the atmosphere and ocean~\cite{boffetta2012two,bouchet2012statistical}. Such streams are defined by the constant horizontal velocities: $u(x,y) = -U/2$ in the upper half-plane $y > 0$ and $u(x,y) = U/2$ in the lower half-plane $y < 0$; the vertical speed $v(x,y)$ is everywhere zero. This flow can be seen as a vortex sheet localized on the horizontal axis $y = 0$, where the vorticity field $\omega = \partial v/\partial x-\partial u/\partial y$ has the Dirac-delta singularity, $\omega = U\delta(y)$.  This is a steady solution to the inviscid incompressible Euler equation.
For the ideal flow, the well-known linear theory~\cite{drazin2002introduction} predicts that a small perturbation with wavenlength $l$ has the exponential temporal growth $\sim \exp(\pi Ut/l)$. The  growth rate becomes infinite when the perturbation scale vanishes, implying the explosive breakdown of  linear theory and  signaling the ill-posedness of the underlying equations~\cite{majda2002vorticity}, as required for  Lorenz's intrinsic unpredictability.
In order to \corr select \rroc  physically relevant solutions, we employ simultaneously two different small-scale mechanisms. 

The first mechanism refers to a vanishingly small perturbation, which can be viewed as the effect of a small noise induced, e.g., by microscopic fluctuations~\cite{ruelle1979microscopic}.   Specifically, we consider initial conditions for the vortex sheet with infinitesimal small-scale perturbations of the form 
\begin{equation}
  \label{eq_IC}
  t = 0:\quad
  \omega_\varepsilon(x,y) = \left[1+\varepsilon \eta(x)\right]U\delta(y),
\end{equation}
where $\varepsilon$ is a small  parameter and $\eta(x)$ is a perturbation profile generated by white noise.

The second mechanism desingularizes the equations of motion at vanishingly small scales, such that the resulting regularized equations can be used to evolve the flow.  
In order to explore the \corr physical properties of the stochastic regularization, \rroc we focus on two fundamentally distinct formulations: (\textit{a}) the viscous or hyperviscous (Navier--Stokes) regularization, which is controlled by the small viscosity parameter $\nu > 0$ and (\textit{b}) the point-vortex (Birkhoff--Rott) approximation,  where the regularization is  determined by the number of vortices $N_b$  and controlled by  the small parameter $\nu \propto N_b^{-1}$. Here, the first case employs a natural dissipative mechanism, and the flow is considered in the so-called Eulerian formulation.
\corr
For deterministic initial data, adding a vanishingly small viscosity is known to select weak solutions
of the 2D Euler equations but is not enough to ensure uniqueness \cite{delort1991existence,majda1993remarks}.
\rroc
 Differently, the second case follows the Lagrangian approach in fluid mechanics,  also relevant for superfluids \cite{johnstone2019evolution,gauthier2019giant},  in which one follows point vortices advected by the induced velocity field.
\corr
Recent numerical studies \cite{suryanarayanan2014free} of the chaotic point-vortex system with finite $N_b$ indicate a type of statistical universality of the non-linear dynamics with respect to the initial data over a wide range of timescales.
While finite amplitude of any random  initial perturbation trivially entails statistical description, robustness of the macroscopic features with respect to the perturbation size points towards a deeper phenomenon.   
In this work we unveil the origin of dynamical universality, showing  that  for vanishingly small values of viscosity and noise, the KH instability develops into a universal and spontaneously stochastic flow. 
\rroc
\begin{figure}
 \includegraphics[width=1\columnwidth]{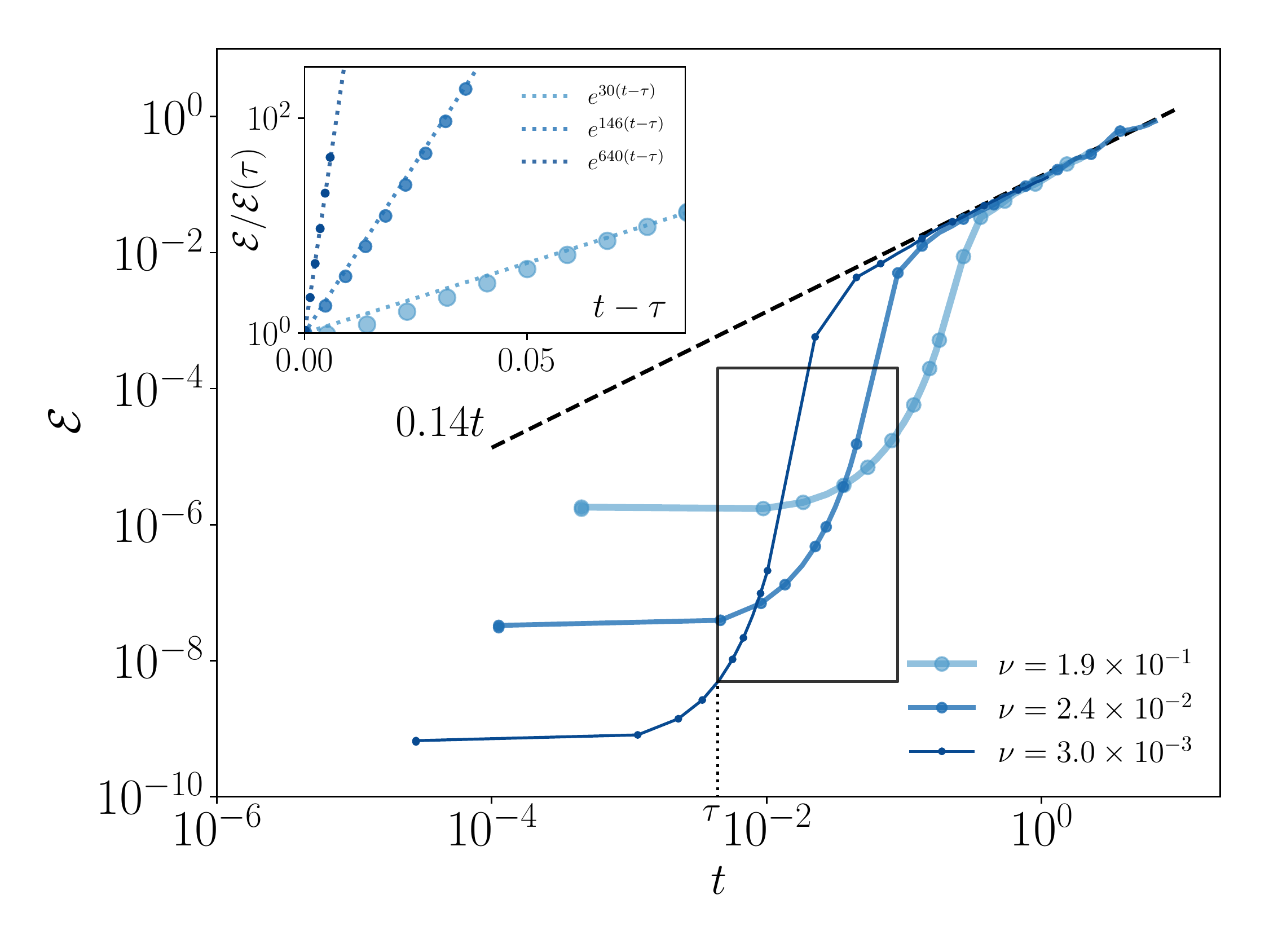}\\
  \caption{Explosive separation between two realizations of the Navier-Stokes  dynamics measured as the energy $\mathcal{E}(t)$ of the difference between two solutions, which are $\varepsilon$-close at the initial time. The values of  viscosity and initial separation corresponding to the three graphs are given in Tab.~\ref{table:param},
  and the observational scale is half the  size of the computational domain $L=\pi$.
The main figure is shown in logarithmic scales and demonstrates the asymptotic algebraic law $\mathcal{E} \approx 0.14 t$ emerging in the limit of vanishing viscosity and initial separation. The inset, shown in linear horizontal and logarithmic vertical scales, zooms into the non-universal initial exponential stage.}
  \label{fig:separationspectrum}
\end{figure}

\begin{figure*}[htb]
      \includegraphics[width=1\textwidth,trim=0cm 0cm 1cm 0cm,clip]{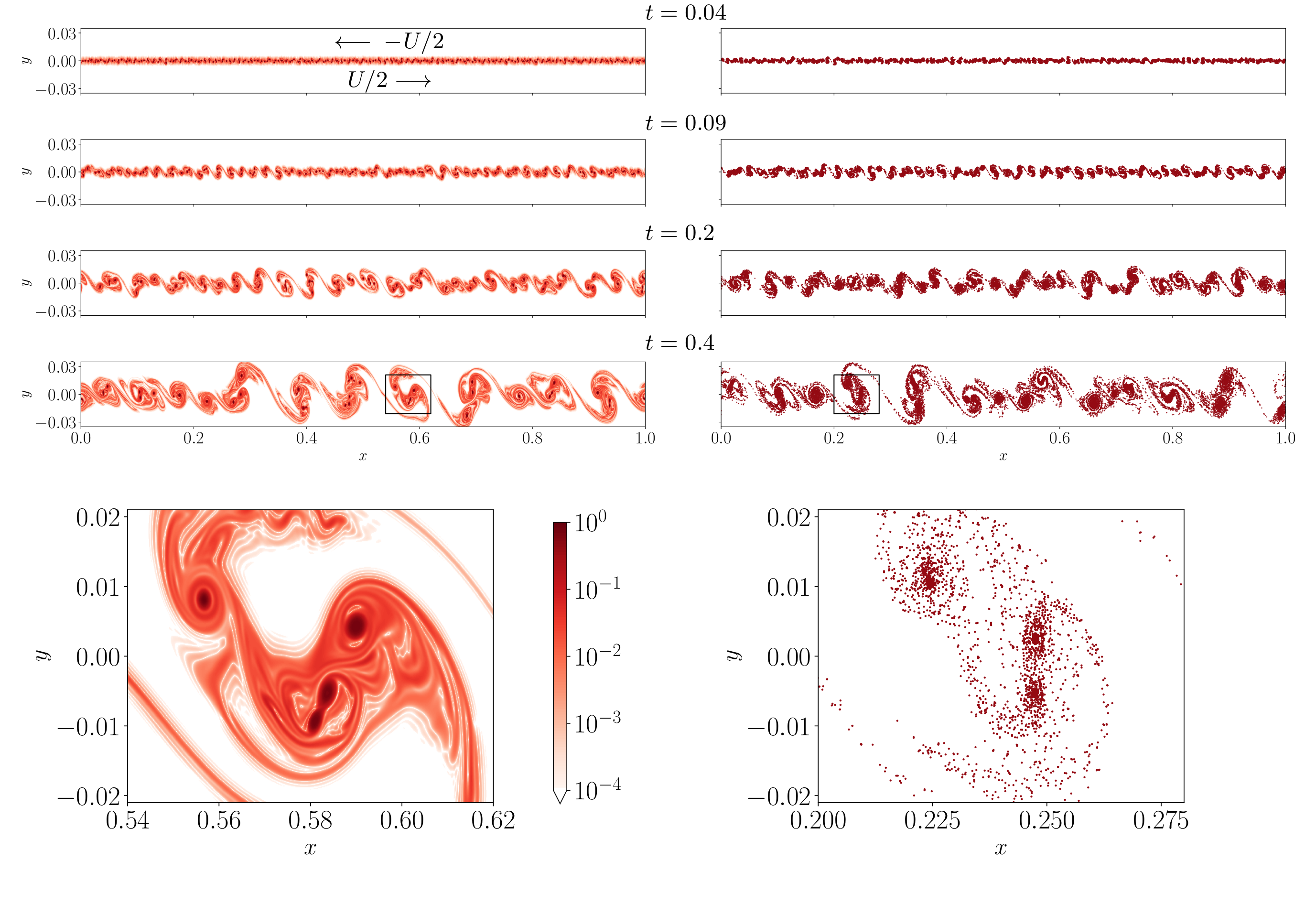}
\caption{\label{fig:snapshots} Snapshots of the vorticity fields at successive times for the  hyperviscous Navier-Stokes dynamics at $\nu=3\times 10^{-3}$ (left) and   Birkhoff--Rott dynamics using $2^{17} = 131,072$ point vortices (right). Bottom panels  zoom in the structures indicated by the rectangular boxes. Space and time variables in all figures are presented in dimensionless form conforming to the velocity $U$ and the size of  computational domain; see the Methods section.}
  \label{fig1}
\end{figure*}

By \textit{spontaneous stochasticity}, we mean that such a flow is intrinsically random. \textit{Universality} signifies  that the statistical properties of the flow are independent of the type of regularization, except at very small scales and times. %
The mathematical formulation of this phenomenon corresponds to the vanishing viscosity and noise, $\nu \to 0$ and $\varepsilon \to 0$, where the two limits are taken simultaneously 
in a suitable fashion, such that physically  meaningful norms of the initial disturbance (such as  the kinetic energy) vanish; see the Methods section for more details. 
\corr
This limiting procedure grants the originally ill-posed inviscid problem with a well-posed statistical interpretation where the asymptotic flow is unambiguously defined as a non-trivial stochastic process.
Mathematically~\cite{flandoli2011random,drivas2020statistical}, 
one expects that each realization of this limiting process 
\rroc
solves the incompressible Euler equations in a weak sense for the initial ideal vortex sheet, $\omega = U\delta(y)$. Physically, this implies that infinitesimal effects of micro-scale regularization and noise do not select a unique deterministic KH flow, but rather a well-defined statistical solution, with universal non-stationary probability distribution.

\bigskip \noindent\textbf{\small The spontaneous butterfly effect.} 
In order to reveal how the intrinsic randomness in the vortex layer emerges at small times, we first focus on the Navier--Stokes dynamics,  with the  viscous parameter $\nu$ here also specifying the noise amplitude as $\varepsilon \propto \nu$. 
We estimate the random component of the flow by measuring how two solutions, $(u,v)$ and $(u',v')$, which differ initially by a very small random perturbation of size $\varepsilon$, split  in time. This is done by introducing the separation energy
	\begin{equation*}
          \mathcal{E}(t) = \frac{2}{L\,U^2}
        \int_{-L/2}^{L/2} dy \langle (u-u')^2+(v-v ')^2\rangle_{x,\eta},
	\end{equation*}
where $L$ is a prescribed (macroscopic) observation scale and the average $\langle \cdot \rangle_{x,\eta}$ is over both the $x$-direction and the statistical realizations of the initial noise. With this choice of normalization, $\mathcal E$ varies between $0$  and $1$, and those two extremal values are reached when the two solutions,  respectively, coincide or fully decorrelate over the observational window.
 The results are presented in 
 Fig.~\ref{fig:separationspectrum}, where the separation energy is plotted for the three different sets of parameters $\nu$ and~$\varepsilon$.  

Initially the flows are very close and diverge exponentially as seen in the inset of Fig.~\ref{fig:separationspectrum}. This exponential growth is governed by a positive Lyapunov exponent, which is the distinctive feature of the usual butterfly effect~\cite{gleick2011chaos}.  However, it is apparent that the growth rate depends strongly on the viscous parameter $\nu$. In fact, the main part of Fig.~\ref{fig:separationspectrum} demonstrates that the exponential growth is a transient behavior.  The transient region is shifted to smaller and smaller times in the combined limit of infinitesimal viscosity and initial perturbation, $\nu \to 0$ and $\varepsilon \to 0$. 
 At larger times, the separation energy reaches the asymptotics %
 $\mathcal{E} \approx 0.14 t$, which is independent of the initial separation and has the order of the total energy within the flow. %
 This behavior quantifies spontaneous stochasticity for infinitesimal viscosity and initial disturbance: solutions which are initially undistinguishable and deterministic become distinct and random at finite times. \rroc
   It is    analogous to Richardson's relative dispersion in fully developed turbulent flow~\cite{frisch1995turbulence}, where the distance between two Lagrangian fluid elements grows algebraically as $\sim t^{3/2}$, independently of how close they initially are. The explosive separation of Fig.~\ref{fig:separationspectrum} can be seen as the Eulerian counterpart to this phenomenon as it pertains to the full description of the flow. This type of randomness is spontaneous since it builds up in infinitesimal time and requires only infinitesimal perturbations to be revealed.

\medskip \noindent\textbf{\small Time development of the vortex layer.}
We will now argue that while the growth of the vortex layer is an ill-posed dynamics from the  deterministic viewpoint, it becomes well-posed in a  probabilistic sense, meaning that macroscopic statistical features of the flow are independent of the noise and of the type of regularization.
Figure~\ref{fig1} shows a typical vorticity distribution for numerical simulations with the two regularization methods; see the Methods section for details of numerical implementations. 
The  panels on the left correspond to  the viscous regularization based on direct numerical simulations of the Navier--Stokes equations in a square domain with periodic boundary conditions. The panels on the right correspond to simulations of the Birkhoff--Rott dynamics.
Visually, the vorticity distributions obtained for each type of regularization look akin to each other on macroscopic scales. It is only upon zooming into the microscopic structure, that  the distinction between the  viscous  and the point-vortex regularization becomes apparent, as the first is continuous and the second is discrete.

On macroscopic scales, the development of the shear layer consists in a cascade process of collisions and the subsequent mergers of smaller vortex blobs into larger ones; see Fig.~\ref{fig1}. 
Here, the blobs are small concentrated regions of high vorticity, surrounded by halos of lower-vorticity streaks~\cite{krasny1986desingularization,lesieur1988mixing,sommeria1991final,lecoanet2015validated}. In the collision process, a part of the vorticity is scattered away from the blobs, and it may be absorbed by the same or other blobs at later times.
Measurements suggests that the vorticity is divided into two approximately equal parts corresponding to the blobs and to the background flow, while the enstrophy  is mostly carried by the blobs.

We now verify numerically that the vortex layer can be qualified as being universal and spontaneously stochastic, namely that the increase of numerical resolution along with decreasing viscosity and noise yield a unique stochastic solution at finite times.
By the scaling theory \cite{barenblatt2003scaling}, one may expect that statistical properties of the macroscopic  dynamics depend only on the two independent quantities, namely, the velocity jump $U$ and the mixing length $\ell(t)$. The latter describes the width of the vortex layer, and can be conveniently defined as 
\begin{equation}
\ell(t) = \left( \dfrac{1}{U}\int \left \langle y^2 \omega(\bR,t) \right\rangle_{x,\eta}dy\right)^{1/2},
\end{equation}
 where $\bR=(x,y)$. The average $\langle \cdot \rangle_{x,\eta}$ is over both the $x$-direction and the statistical realizations of the initial noise.  For the Birkhoff-Rott regularization, this quantity represents the standard-deviation of point-vortices across the vortex layer.

\begin{figure}[t]
  \includegraphics[width=\columnwidth]{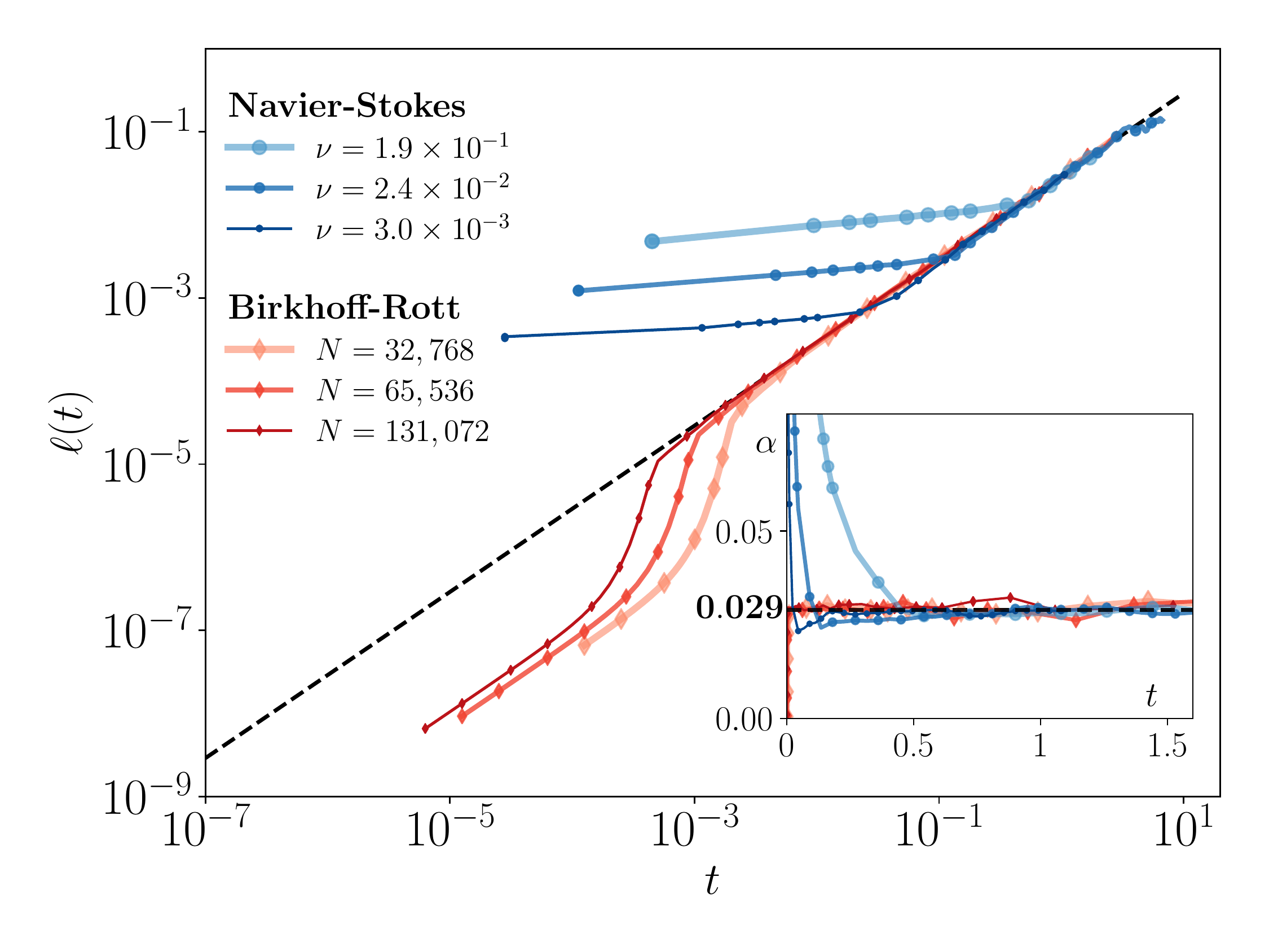}
  \vspace{-18pt}
  \caption{Time evolution for the width $\ell(t)$ of the vortex layer, for both the Navier--Stokes (blue) and  Birkhoff--Rott (red) regularizations. 
Inset shows the asymptotic convergence for  the universal dimensionless pre-factor $\alpha = \ell/Ut \approx 0.029$.
  }
  \label{fig:width} 
\end{figure}

From dimensional analysis, we expect that the mixing layer grows linearly in time: $\ell = \alpha U t$, where $\alpha$ is a dimensionless coefficient.
Figure~\ref{fig:width} presents the numerical results shown in log-log scale, which verify this asymptotic linear scaling law.
The vortex layer is formed after a short transient time, and this transient time decreases as we decrease the regularization parameter and noise. The subsequent evolution yields the universal pre-factor $\alpha \approx 0.029$ for both Navier--Stokes and Birkhoff--Rott regularizations; see the inset in Fig.~\ref{fig:width}.

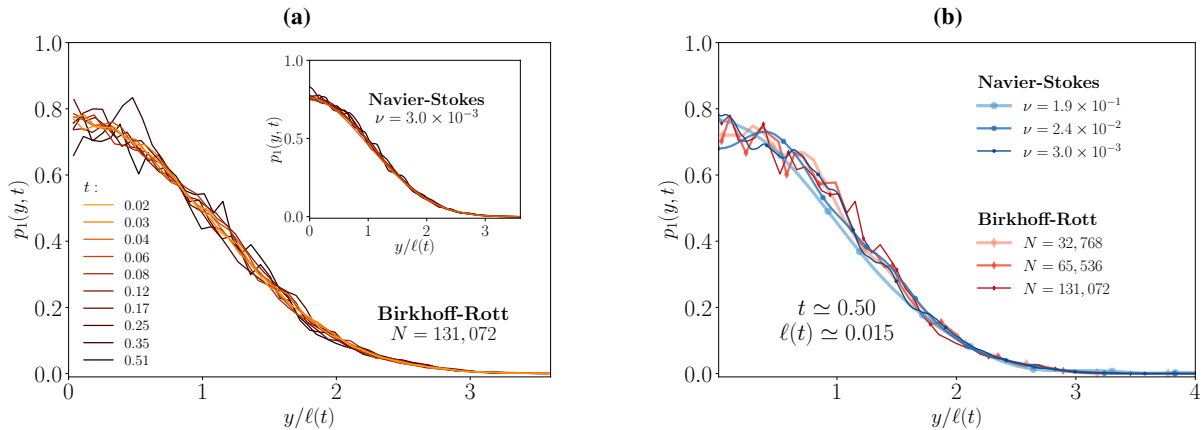
\begin{figure*}
  \includegraphics[width=0.9\columnwidth]{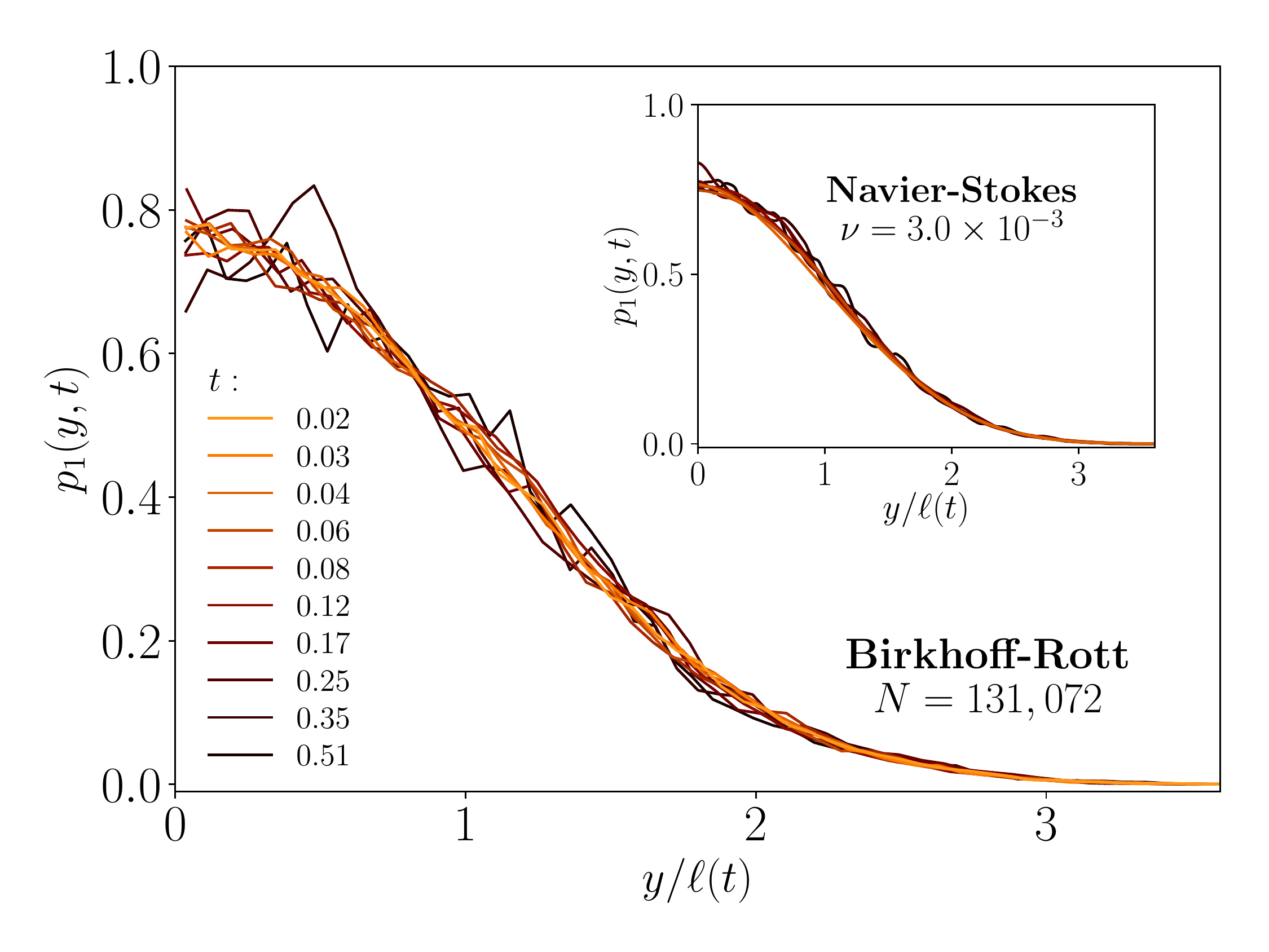}
  \hspace{20pt}
  \includegraphics[width=0.9\columnwidth]{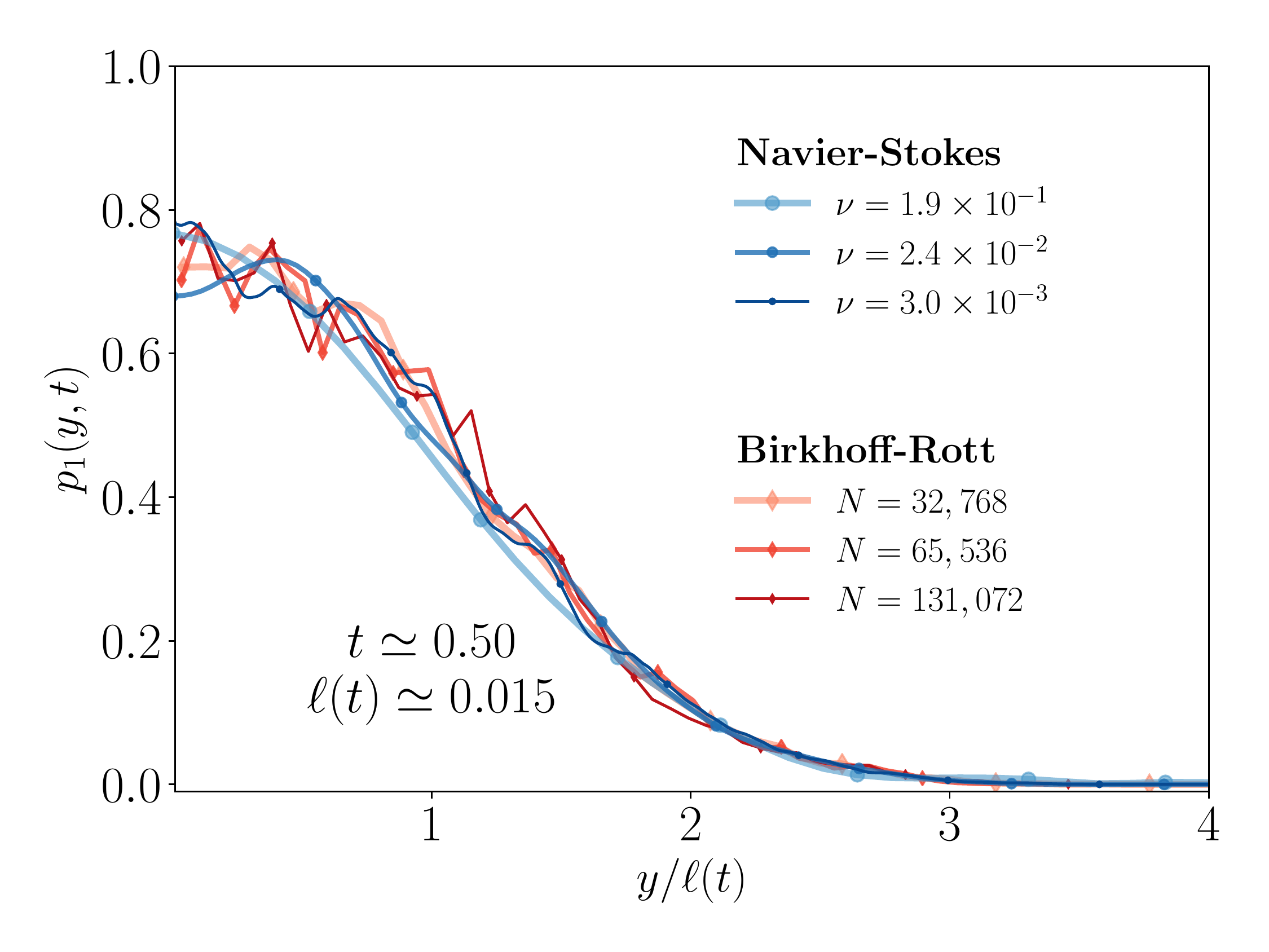}
  \vspace{-35pt}
  \setlength{\unitlength}{\columnwidth}
  \begin{picture}(1.0,0.1)(0,0)
    \put(0.0,0.76){{\bf(a)}}
    \put(1,0.76){{\bf(b)}}
  \end{picture}
  \caption{(a) Vorticity profile $p_1(y,t)$ within the mixing layer as a function of the rescaled variable $y/\ell(t)$ at various times logarithmically spaced between $t=0.02$ and $t=0.5$. (b) The same graph, now at fixed time $t=0.5$ for both Navier--Stokes and the point vortex simulations.}
  \label{fig:profile} 
\end{figure*}

\medskip
\noindent\textbf{\small Statistical universality.}
We now study multi-scale statistical properties of the flow. 
We start by analyzing the evolution of the normalized vorticity profile, 
\begin{equation}
	  p_1(y,t) = \dfrac{\ell(t)}{U}\left\langle \omega (\bR,t)\right\rangle_{x,\eta}, 
 \end{equation}
obtained from the vorticity distribution averaged with respect to the $x$-direction and the statistical realizations.
In the Birkhoff-Rott case, $p_1$ relates to the one-point  distribution of point-vortices along the $y$ direction so that $\int  p_1 (y,t)\;dy =\ell(t)$.
Numerical vorticity profiles are shown in Fig.~\ref{fig:profile}. Panel (a)  highlights  the self-similarity of this one-point statistical quantity, that is $p_1(y,t) \approx P_1\left(y/\ell(t)\right)$. Panel (b) confirms that the asymptotic profile $P_1$, corresponding to vanishing amplitudes of the initial perturbations, is identical for both the Navier--Stokes and Birkhoff--Rott regularizations.

As a second-order statistical observable, we introduce the isotropic two-point correlator 
\begin{equation} 
    p_2 (r,t) =  \frac{1}{U^2} \int   
\big\langle \omega(\bR,t)\omega(\bR + \boldsymbol{\rho},t) \big\rangle_{x,\eta}   
\delta\left( \vn{\boldsymbol{\rho}} -r\right)
d^2\!\boldsymbol{\rho} \, dy
  \label{eq_p2}
\end{equation}
written in dimensionless form.
It is constructed from the covariance of the vorticity field between the points $\bR=(x,y)$ and $\bR+\boldsymbol{\rho}=(x+\rho_x,y+\rho_y)$ at a given distance $\vn{\boldsymbol{\rho}} = r$. 
The function $p_2 (r,t)$ is strongly dominated by the relative positions of concentrated blobs of vorticity (see Fig.~\ref{fig1}), for which the integrated product of vorticities in (\ref{eq_p2}) is maximal. 

\begin{figure*}
  \vspace{8pt}
  \includegraphics[width=0.95\columnwidth]{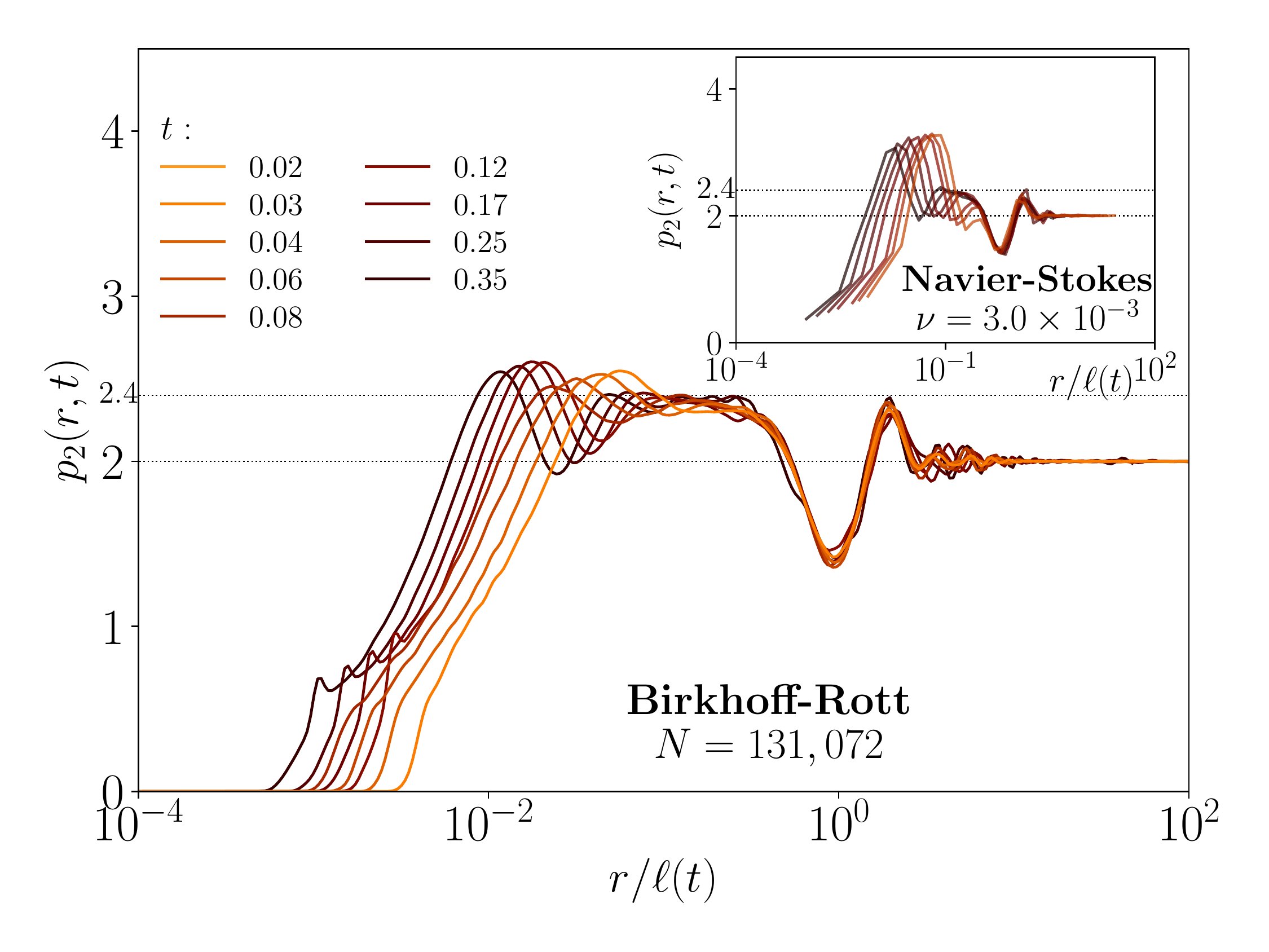}
  \hspace{20pt}
  \includegraphics[width=0.95\columnwidth]{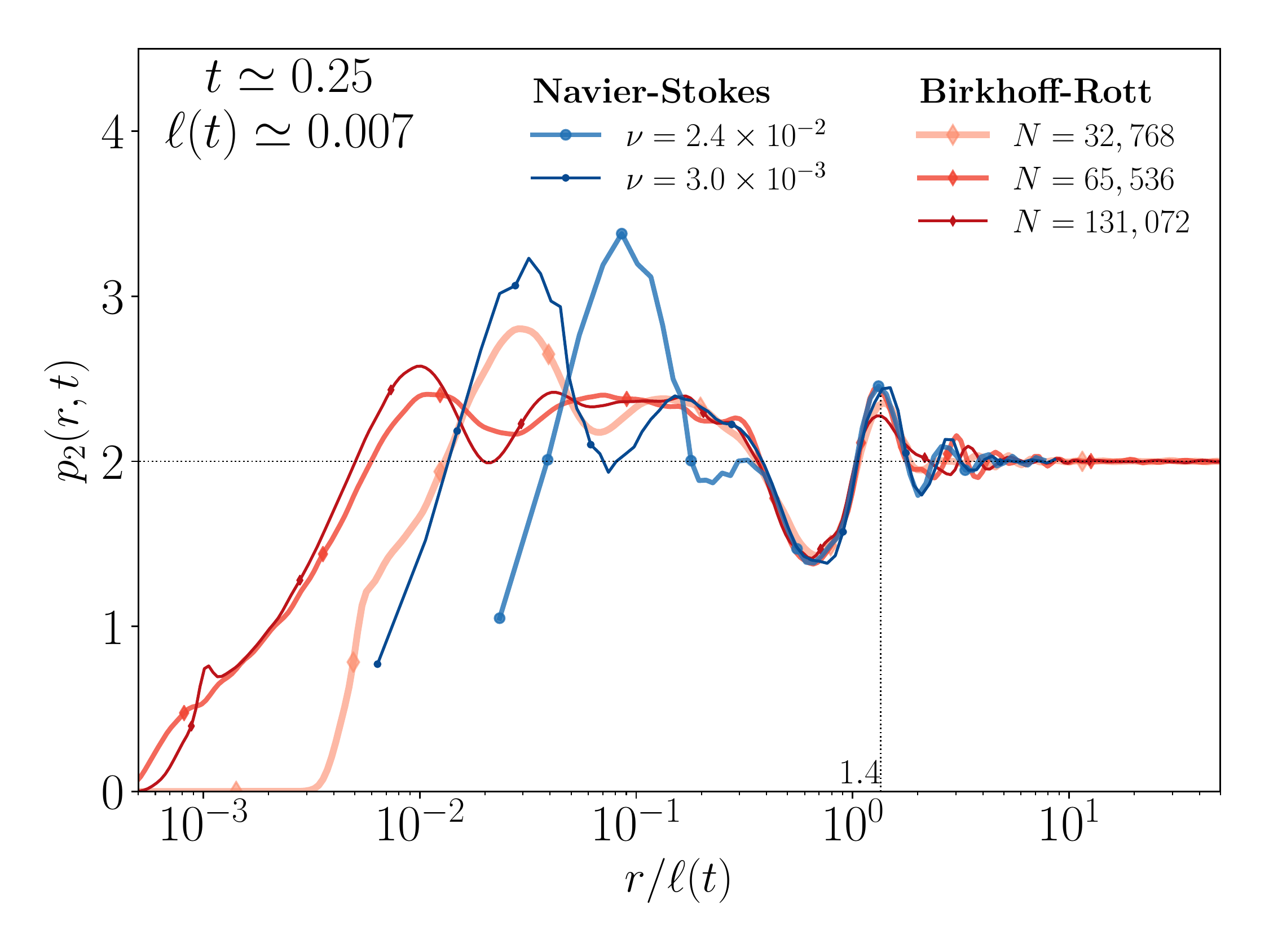}
  \vspace{-35pt}
  \setlength{\unitlength}{\columnwidth}
  \begin{picture}(1.0,0.1)(0,0)
    \put(0.0,0.8){{\bf(a)}}
    \put(1,0.8){{\bf(b)}}
  \end{picture}
  \caption{(a) The function $p_2(r,t)$ measuring two-point correlation properties at various times logarithmically spaced between $t=0.02$ and $t=0.5$. The main figure corresponds to the point-vortex simulation, while the inset shows the corresponding data for the hyperviscous run.
    (b) The same function at the fixed time $t=0.25$ for both Navier--Stokes and Birkhoff--Rott simulations. }
  \label{fig:p2}
\end{figure*}
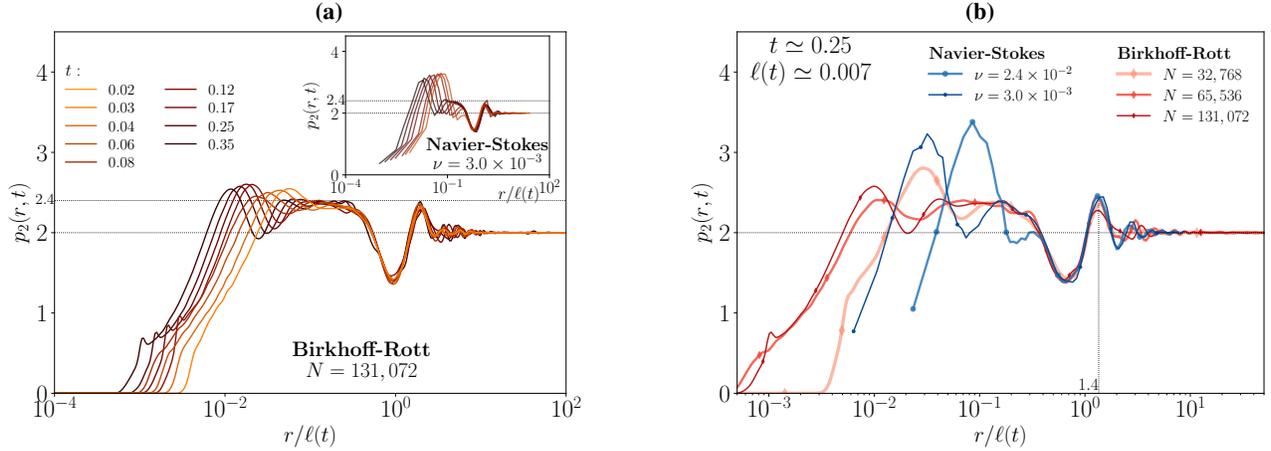

Results of numerical simulations are shown in Fig.~\ref{fig:p2}, where Panel (a) displays the asymptotic self-similar relation $p_2(r,t) \approx P_2\left(r/\ell(t)\right)$.
Panel (b) demonstrates that the asymptotic form of $p_2$ is universal: at a given finite time, it collapses onto a single universal function $P_2$ in the limit of vanishing regularization and noise for both Navier--Stokes and Birkhoff--Rott formulations. One can see that the profiles for different regularizations are distinct only at small scales, which reflects the different small-scale structures visible in the bottom plots of Fig.~\ref{fig1}. The function $P_2$ has a pronounced maximum around $r = 1.4\,\ell(t)$ characterizing a typical distance between nearest vortex blobs in Fig.~\ref{fig1}. 
For very large distances $r \gg \ell(t)$, it converges to the limiting value $P_2 \to 2$. This value can be computed by substituting the  vortex sheet expression $\omega = U\delta(y)$ into (\ref{eq_p2}), which approximates the vortex layer at large scales. It is remarkable that a constant state develops also at small distances $r \ll \ell(t)$, with the universal value estimated as $P_2 \to 2.4$; see Fig.~\ref{fig:p2}. This value characterizes the uniform statistical distribution of the vorticity at distances which are small compared to the mixing length, but still larger than the regularization scale. In our view,  this constant asymptotic value reflects the self-similar nature of the reconnection process,  where vortex blobs constantly attract each other, before merging into larger blobs.

\bigskip
\noindent\textbf{Discussion}

\noindent Opposed to the exponential growth of observational errors in chaotic systems, now known as the butterfly effect, Lorenz anticipated that finite-time  randomness can emerge in some formally deterministic multi-scale fluid systems irrespective of how small the initial uncertainty~\cite{lorenz1969predictability}. In our work, we demonstrated that this scenario is relevant for the flows generated by the Kelvin--Helmholtz instability of a singular shear layer. Being ill-posed for the ideal fluid mechanics, this system triggers a \emph{spontaneous stochastic process} for the vorticity field, when considered in the combined limit of vanishing micro-scale regularization and vanishing random perturbation. In particular, considering hyperviscous Navier--Stokes \textit{vs.}\/ point-vortex Birkhoff--Rott formulations, we showed that this limiting stochastic process is universal. 

The recently developed theory for particles advected by non-smooth deterministic velocity fields~\cite{eijnden2000generalized,gawedzki2001turbulent,falkovich2001particles,le2002integration,kupiainen2003nondeterministic} demonstrate similar spontaneously stochastic properties when infinitesimal noise is introduced: fluid parcels  separate in finite time  no matter how close they are initially. Such an approach proved to be relevant for particles in turbulent flows~\cite{jullien1999richardson,scatamacchia2012extreme,bitane2013geometry,bitane2012time,buaria2015characteristics}, where it is referred to as the Richardson super-diffusion~\cite{frisch1995turbulence}.  Within this modern perception, Lorenz's scenario of intrinsic randomness can be seen as the Eulerian counterpart of spontaneous stochasticity applied to the flow \textit{fields} rather than to individual particles\cite{mailybaev2016spontaneously}. In our point of view, this extension is rather natural, because the Eulerian description of fluid dynamics involves  the transport of active quantities, such as momentum or vorticity. {\color{black}A comprehensive mathematical framework of this effect dealing with statistical solutions to fluid equations\cite{fjordholm2016computation,fjordholm2017statistical}, however, is still to be developed~\cite{flandoli2011random}.}

Whether the finite-time unpredictability is caused by a small though finite ``butterfly'' as in chaos or by an arbitrarily small perturbation as in spontaneous stochasticity, the distinction may be illusive for an unprepared observer. However, these two phenomena represent fundamentally different physical and mathematical mechanisms that bridge the deterministic classical world to genuinely random systems. As such, spontaneous stochasticity could then possibly provide the salient mechanism that generates  intrinsic  randomness in natural phenomena. \corr It provides a conceptual framework that justifies the relevance of stochastic modeling for apparently deterministic  fluid dynamical systems \cite{palmer2019stochastic}, as well as for systematic statistical mechanics  approaches \cite{suryanarayanan2014free}\rroc. In addition to potential applications in fluid mechanics, which of course include the open problem of developed turbulence \cite{brenner2016potential} or  the physics of boundary layers\cite{drivas2017lagrangian,waidmann2018energy}, the relevance of the spontaneous stochasticity phenomenon may extend to other nonlinear field theories featuring multi-scale dynamics and singularities, e.g., wave turbulence\cite{nazarenko2011wave},  nonlinear optics \cite{kraych2019statistical} and astrophysics~\cite{genel2019quantification,khalatnikov1985stochasticity}.

\medskip
\noindent\textbf{Methods}

\begin{footnotesize}

\noindent\textbf{Navier--Stokes regularization.}
Our simulations were performed using the open-source GHOST parallel solver \cite{gomez2005parallel,mininni2011hybrid}, that employs a pseudo-spectral scheme with standard 2/3 dealiasing on spatial grids with $N_a^2$ points, and second-order Runge--Kutta scheme in time.  
The timestep is fixed and characterized by  Courant number $\sim\!0.2$. 
Specifically, the Navier--Stokes equations in vorticity form
\begin{equation}
	\partial_t \omega + \bu \cdot \nabla \omega = D_\nu \;\;\text{with}\;\; {\bf u}  = -\nabla^\perp \left(\Delta^{-1} \omega\right)
\end{equation}
are integrated within a two-dimensional periodic domain $(x,y) \in [0,\,2\pi]^2$. 
Results presented in the paper are obtained using a hyperviscous linear dissipation   defined through its Fourier-space representation as $\widehat{D}_\nu  =- \nu k^2(k/k_{\max})^6\widehat{\omega}$,  where the regularization wavenumber $k_{\max}=N_a/3$ is used; this is a common tool to localize viscous effects at small scales. Simulations with the standard viscous dissipation $D_\nu = \nu\Delta \omega$ were also carried out employing up to $16, 384^2$ grid points: they lead to the same conclusions though for a shorter interval of scales due to numerical limitations.

To comply with periodicity, the initial conditions (\ref{eq_IC}) are implemented along two parallel lines: $y = 0$ and $y = \pi$, so that the total vorticity is zero. The noise $\eta(x)$ is generated by a centered uniform stochastic process with spatial correlations $\left\langle \eta(x)\eta(x^\prime)\right\rangle=\delta(x-x^\prime)/3$. 
Simulations are stopped at sufficiently small times to avoid interaction between the layers and nonlocal effects of a finite domain. Then, the two resulting vortex layers are analyzed separately in the respective domains. 

We performed nine simulations for each of the three parameter sets from Tab.~\ref{table:param}. 
These parameters follow the rule $k_{\max} \propto N_a$, $\nu \propto N_a^{-3/2}$ and $\varepsilon \propto N_a^{-3/2}$, so that $\nu\to 0$ and $\varepsilon \to 0$ simultaneously with the increasing resolution $N_a$. 
 To avoid spurious Gibbs oscillations of rough initial conditions, 
the initial state is diffused by hyperviscosity  until  the largest-wavenumber  $k_{\max}$ has negligeable Fourier contribution, comparable to the machine precision.
This takes a short initialization time $t_d \propto N_a^{-1/2}$, vanishing with increasing resolution. 
After this initial smoothing, the energy and enstrophy (squared norm of the vorticity) of the initial perturbation  in expression (\ref{eq_IC})  scale as 
$\mathcal E(0^+) \sim N_a^{-3}\log N_a$ and $\mathcal Z(0^+) \sim N_a^{-1}$:  Both quantities vanish upon increasing resolution.
The initial (regularized) conditions therefore converge to the straight (unperturbed) vortex sheet in both energy and enstrophy metrics. We observe that  our choice of scaling guarantees that the most unstable linear mode in our simulation is $\propto\!N_a$ upon increasing resolution.  

\corr Finally, please note that choosing 
simultaneous scaling rules for $\varepsilon \to 0$ and $\nu\to0$ 
is dictated by the fact that these limits in principle do not commute. In particular, letting $\varepsilon \to 0$ before $\nu \to 0$, one should expect the process to converge towards the  deterministic trivial base flow. 
\rroc

 \begin{table}[ht]
  \begin{tabular}{|cccc|}
    \hline
    $N_a$  & $k_{max}$  & $\nu$ & $\varepsilon$ \\
    \hline
    $512$  & $170$  &  $1.9\times 10^{-1}$ & $2.2\times10^{-3}$ \\
    $2048$  & $682$  & $2.4\times10^{-2}$ & $2.8\times10^{-4}$ \\
    $8192$  & $2730$  & $3.0\times10^{-3}$ & $3.5\times10^{-5}$ \\
    \hline
  \end{tabular}
  \caption{ Parameters for the hyperviscous Navier--Stokes simulations with $U = 1/\sqrt 2$.}
  \label{table:param}
\end{table}

\noindent\textbf{Birkhoff--Rott regularization.} 
In this case, we use a domain that is $1$-periodic in the $x$-direction and unlimited in the $y$-direction. The initial condition (\ref{eq_IC}) is approximated by a periodic discrete row of point-vortices located on the line $y = 0$ at positions $x_n(0) = n/N_b$, $n = 1,\ldots,N_b$, and carrying the initial vorticity $\omega_n = \frac{U}{N_b}\delta(x-x_n)\delta(y)(1+\varepsilon\, N_b^{1/2}\,\eta_n)$, where the $\eta_n$'s are independent Gaussian variables with unit variance.  Using the complex position variables $z_n =x_n + i y_n$, Biot-Savart law prescribes the advection of each point-vortex as\cite{aref2007point,krasny1986desingularization} 

\begin{equation}
	\dot z_n^\star = \dfrac{1}{2i}\sum_{\substack{1 \le m \le N_b \\ m \neq n}} \omega_m\,\cot\left[ \pi\,(z_n-z_m)\ \right],
	\label{eq:PV}
      \end{equation}
      where the star subscript denotes complex conjugation.
      We performed simulations of the dynamics (\ref{eq:PV}) with $U=1$, $\varepsilon=10^{-5}$, and $N_b = 2^{15}$, $2^{16}$ and $2^{17}$ vortices using the  fourth-order Runge--Kutta scheme. The timestep $\Delta t$ is chosen adaptively to satisfy
 $\Delta t~<~0.1\,\min_{n\neq m} \left\lbrace|z_n-z_m|/ |\dot{z}_n-\dot{z}_m|\right\rbrace$.
This sort of Courant-Friedrichs-Lewy upper-bound is enough to ensure stability and accuracy of the numerical scheme. Indeed, with such a choice, the kinetic energy, defined as
        \begin{equation}
	E = -\dfrac{1}{4\pi}\sum_{\substack{1 \le n,m \le N_b \\ n \neq m}} \omega_n\,\omega_m\,\log\left|\sin\left[ \pi\,(z_n-z_m)\ \right]\right|,
	\label{eq:KE_PV}
\end{equation}
is conserved with a high-enough accuracy. At the end of the simulations, the relative error in energy remains below $10^{-4}$. 
\corr
Qualitatively similar CFL criteria were employed in  previous simulations of point-vortex systems in similar set-ups
\cite{aref1980vortex,suryanarayanan2014free}. 
While this CFL criterion in principle does not preclude divergence of dynamical trajectories due to numerical noise, it is enough to ensure convergence in a statistical sense. Upon increasing resolutions, the space-time distributions of point-vortices indeed converge towards a non-trivial measure. In mathematical terms, the observed  statistical universality reflects convergence in distribution \cite{jacod2012probability}.
\rroc

\noindent\textbf{Finite-size effects.}
Formally,  the scaling theory for  the singular vortex sheet addresses flows evolving within an  unbounded domain. In our numerical simulations, this domain has a finite 
size in the $x$-direction, taken as $L=2\pi$ for the Navier--Stokes and $L=1$ for the Birkhoff-Rott models. In all figures, we displayed the results in dimensionless form using $t_* = L/U$ as the time unit, for which 
the mixing layer invades  roughly  $3\%$ of the computational domain. 

\end{footnotesize}

\medskip
\noindent\textbf{Data availability}

\begin{footnotesize}
  \noindent The data that support the findings of this study are available from the corresponding author on request.
\end{footnotesize}

\medskip
\noindent\textbf{Code availability}

\begin{footnotesize}
  \noindent The simulation and post-processing codes that have been used to produce the results of this study are available from the corresponding author on request.
\end{footnotesize}

\medskip
\noindent\textbf{Acknowledgments}
\begin{footnotesize}
  \noindent We thank  M. Brachet and P. Mininni for insightful discussions and G. Eyink for sharing his continuously encouraging  
perspectives on the phenomenon of spontaneous stochasticity. This work was supported by CNPq (grants 303047/2018-6, 406431/2018-3), Petrobras-SIGITEC (grant 2018/00650-9) and the Brazilian-French Network in Mathematics.
\end{footnotesize}

\bibliography{biblio}

\end{document}